\documentclass[12pt,reprint]{revtex4-1}

\usepackage{graphicx}
\usepackage{url}

\begin{document}

\title{Optically-dressed resonant Auger processes induced by high-intensity x rays}

\author{Antonio Pic\'on}
\email{apicon@anl.gov}
\affiliation{Argonne National Laboratory, Argonne, Illinois 60439, USA}
\author{Phay J. Ho}
\affiliation{Argonne National Laboratory, Argonne, Illinois 60439, USA}
\author{Gilles Doumy}
\affiliation{Argonne National Laboratory, Argonne, Illinois 60439, USA}
\author{Stephen H. Southworth}
\affiliation{Argonne National Laboratory, Argonne, Illinois 60439, USA}

\date{\today}

\begin{abstract}
We have unveiled coherent multiphoton interferences originating from different quantum paths taken by the Auger electron induced by a high-intensity x-ray/XUV pulse under the presence of a strong optical field. These interferences give rise to a clear signature in the angle-resolved Auger electron spectrum: an asymmetry with respect to the energy of the Auger decay channel. In order to illustrate this effect we have considered the resonant Auger decay of the transition $2p^{5} \!\leftrightarrow\! 1s^{-1}2p^{6}$ in Ne$^{+}$.  The simulations show that these interferences are very sensitive to the parameters of the x-ray and optical fields.
\end{abstract}

\maketitle


\section{Introduction}

Free-electron lasers (FELs) can achieve very high intensities (more than 10$^{15}$ W/cm$^2$), an important feature to produce nonlinear processes in the x-ray/XUV regime \cite{Linda2010}. The nonlinearity may arise from perturbative sequential absorption \cite{Nina2007,Linda2010,Fritzsche2009}, perturbative nonsequential multiphoton interaction \cite{Makris2009,Meyer2010,Gilles2011,Hishikawa2011}, or nonpertubative interaction \cite{Nina2008,Kanter2011,Nina2012,Cavaletto2012,Demekhin2011,Nikolopoulos2011,Demekhin2012}.  Self-seeding and optical-laser-seeding methods are being developed at FELs that produce x-ray/XUV pulses with high temporal coherence \cite{Amann2012,Allaria2012}.  The combination of high intensity and high coherence enables population control via Rabi oscillations.

Also, the combination of x-ray/XUV light with strong optical fields (10$^{10}$-10$^{15} $W/cm$^2$) introduces a new degree of controllability exceptionally beneficial for pump-probe experiments \cite{Drescher2002,Predrag2011,Bryan2012}, optical control of x-ray absorption \cite{Glover2009,Buth2007}, and x-ray pulse characterization \cite{Dusterer2011}. In these experiments inner-shell holes are created and the Auger decay that follows provides an ultrafast internal probe of the electron dynamics. However, the Auger electron is streaked by the optical field during the electron emission and a proper understanding of the light-matter interaction is needed to analyze the spectrum. In particular, when the Auger electron emission occurs during a significant change of the optical vector potential, the electron wavepacket interferences give rise to a multipeak structure in the spectrum; the so-called sidebands \cite{Schins1994,Smirnova2003,Drescher2005,Kazansky2009,Buth2009,Kazansky2010,Meyer2012}. The sidebands are related to the above-threshold ionization (ATI) phenomenon \cite{Agostini1979}, and analogously to ATI, every sideband-peak is separated by an optical photon energy. Sidebands were identified as essential features in recent work \cite{Picon2013} in which we studied resonant Auger processes under strong optical fields that couple core-excited states. Resonant Auger processes are especially interesting as site-specific excitations {\cite{ReviewMehlhorn,ReviewArmen,ReviewPiancastelli}. However, none of the previous works considered x-rays intense enough to observe nonlinear effects in the sidebands of the Auger electron spectrum.

In the present work we describe a general effect that can be observed in the Auger electron spectrum when a high-intensity x-ray/XUV pulse is used to resonantly excite an inner-shell or core-shell electron under the presence of a strong optical field. This effect can be explained from the energy domain picture or from the time domain picture. From the energy domain picture, the nonlinear effects induced by the intense x-ray/XUV pulse yield an overlapping between consecutive sideband-peaks. The overlapping can be understood as the interferences of different quantum paths taken by the Auger electron. Because of the phases between the consecutive sideband-peaks, their interferences can produce asymmetries with respect to the energy and angle of Auger electron emission. From the time domain picture, the high-intensity and coherent x rays control the inner- or core-shell state dynamics and therefore control when Auger decay occurs. In particular, allowing Auger decay only during the maxima of the optical vector potential, we observe a strong optical streaking that explains the above mentioned asymmetries in the Auger electron spectrum.

In order to demonstrate and explain this effect, we have considered the resonant Auger decay of the transition $2p^{5} \!\leftrightarrow\! 1s^{-1}2p^{6}$ in Ne$^{+}$. Our simulations show that these interferences are very sensitive to the x-ray and optical parameters due to the intrinsic coherence of the whole process.


\section{Theoretical model} \label{sec:theoretical_model}

\begin{figure}[b] 
\centerline{\includegraphics[width=1.\columnwidth,clip]{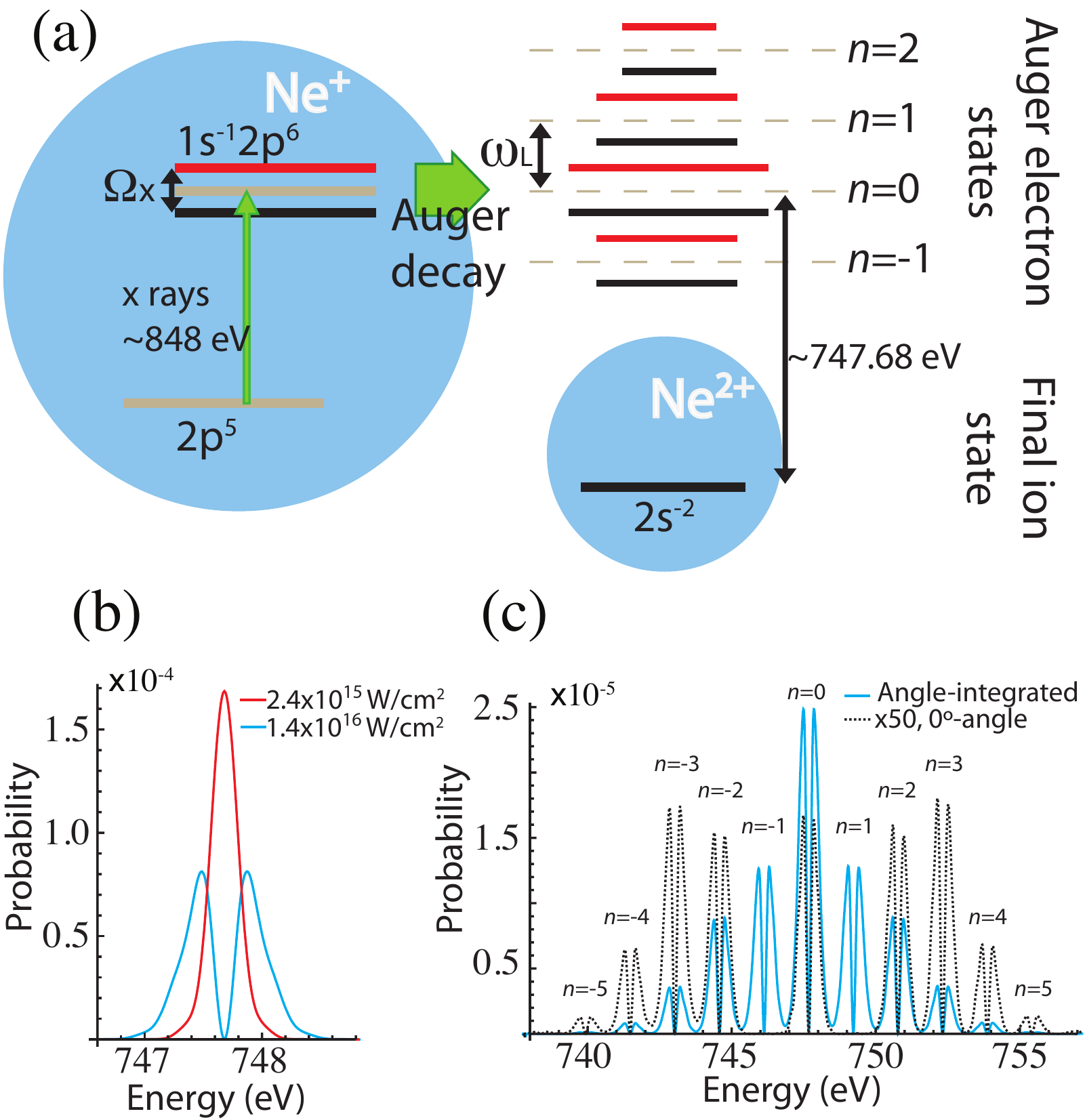}} 
\caption{(Color online)
(a) Sketch of the physical scenario that we consider with $\Omega_x$ the Rabi frequency of the $2p^{5} \!\leftrightarrow\! 1s^{-1}2p^{6}$ transition and $\omega_L$ the optical laser frequency.  (b) Angle-integrated Auger electron spectrum of the resonant transition $2p^{5} \!\leftrightarrow\! 1s^{-1}2p^{6}$ decaying to the final ion $2s^{-2}$($^1$S) for two different intensities of the x rays: $2.4\times10^{15}$ W/cm$^2$ and $1.4\times10^{16}$ W/cm$^2$. (c) Same as (b) with the intensity of the x rays being $1.4\times10^{16}$ W/cm$^2$, but now the continuum is dressed by a strong optical field (wavelength 800 nm and intensity $10^{11}$ W/cm$^2$). The continuous (blue) line is the angle-integrated Auger electron spectrum. The dash (black) line is the Auger electron spectrum in the direction $\theta=0$ (parallel to the polarizations of the x rays and the optical field).
}\label{wavelength} \label{Fig_1}
\end{figure}

The novel phenomenon presented in this work can be extended to any inner-shell or core-excited state interacting with a high-intensity x-ray/XUV pulse, but in order to show some realistic parameters we will present a model for neon, in particular the $2p^{5} \!\leftrightarrow\! 1s^{-1}2p^{6}$ transition in Ne$^{+}$. Intense x-ray effects on the resonant Auger electron lineshape of this transition have been studied experimentally and theoretically \cite{Kanter2011,Nina2012}, and the resonance fluorescence spectrum has been calculated as a two-level system \cite{Cavaletto2012}. Unlike Ref. \cite{Nina2012}, we do not consider the decoherence introduced in the system by using the x rays to both ionize the Ne 2p electron and induce the $2p^{5} \!\leftrightarrow\! 1s^{-1}2p^{6}$ transition in Ne$^{+}$.  For simplicity, we assume that the initial state is the ground state of Ne$^{+}$. In our theoretical model for Ne$^{+}$, see appendices for a detailed description, the x rays couple resonantly the ion ground state ($\vert 0 \rangle$) with the Rydberg core-excited state $1s^{-1}$ ($\vert 1 \rangle$), see Fig. \ref{Fig_1}(a). The intensity of the x rays is strong enough to induce Rabi oscillations between the core-excited state and the ground state. Although the Rabi cycling is the main dynamics induced by the x rays, we include the possibility of ionizing an electron from both the ground and the core-excited state. The core-excited state has a finite lifetime of 2.4 fs and it will decay and emit an Auger electron during the Rabi cycling.  Since our system is exposed to a strong optical field, we account for the fact that the continuum of the Auger decay is dressed by the optical field.  This dressed continuum gives rise to a multipeak structure in the Auger electron spectrum, the sidebands \cite{Schins1994,Smirnova2003,Drescher2002,Drescher2005,Kazansky2009,Buth2009,Kazansky2010,Meyer2012,Picon2013}. In our previous study \cite{Picon2013}, we considered the continuum states of the Auger electron and the optically-induced continuum-continuum transitions responsible for sidebands. The core-excited state may undergo resonant Auger decay into the states $\vert {\bf v}_a, i \rangle$, where $i$ represents the state of the final ion left after the decay with an Auger electron whose velocity is ${\bf v}_a$. We derive the equations of motion (EOM) in the dipole and Markov approximations (or Wigner-Weisskopf theory) \cite{Wigner30,Knight90,Buth2009}


\begin{eqnarray}
&& \mathrm{i}\, {\dot a}_{0}(t) = E_0 a_0(t) + \mu_{01} \varepsilon_{x} (t) a_1(t) \; , \nonumber \\
&& \mathrm{i}\, {\dot a}_{1}(t) = [E_1  -\mathrm{i}{\Gamma_1\over2}] a_1(t) + \mu_{10} \varepsilon_{x} (t) a_0(t)  \; , \nonumber \\
&& \mathrm{i}\, \dot{b}_i ({\bf v}_a,t) =  E_i({\bf v}_a) b_i({\bf v}_a, t) + \gamma_{i,1}({\bf v}_a) a_1 (t)  \nonumber \\ 
&& +\! \int \!\! d{\bf v}_a'^{3} \; \mu_{i,i}({\bf v}_a;{\bf v}'_a) \varepsilon_L (t) b_i ({\bf v}'_{a}, t) \; ,  \label{EOM}
\end{eqnarray}
where $a_0(t)$, $a_1(t)$, and $b_i({\bf v}_a, t)$ represent the amplitudes of the ion ground state, the core-excited state $1s^{-1}2p^{6}$, and the final ionic state $i$ with the Auger electron respectively. Note that atomic units are used throughout. We assume Gaussian pulses for the x rays and a trapezoidal pulse shape for the optical field, see appendix \ref{app:1}. The energies of the ground and core-excited states are given by $E_{k}$, and for continuum states by $E_i({\bf v}_a)={\bf v}_a^2/2 + E_i^{+}$, where ${\bf v}_a^2/2$ is the Auger electron kinetic energy and $E_i^{+}$ is the final ion energy. The dipole moment from state $a$ to $b$ is $\mu_{ba}$. The total decay $\Gamma_k$ accounts for the decay of the core-excited states due to non-radiative (e.g. Auger) and radiative processes as well as of the ionization processes by the optical laser and the x rays \cite{SchmidtBook,Coreno99,Picon2013}. At $10^{11}$ W/cm$^2$, the optical field is intense enough to dress the continuum but does not ionize the system. The transition matrix elements $\gamma_{i,k}({\bf v}_a) \!=\! \langle {\bf v}_a,i\vert \hat{V}_{ee} \vert k \rangle$, $\hat{V}_{ee}$ being the electron-electron Coulomb interaction \cite{Buth2009}, do not depend on time, and within the Wigner-Weisskopf and the dipole approximations we can derive (see more details in Ref. \cite{Picon2013})
\begin{eqnarray} \label{Auger_moment}
\gamma_{i,k}({\bf v}_a) = {e^{i\xi_i^{(k)}} \over\sqrt{(4\pi \vert {\bf \tilde{v}}_a\vert)}} \sqrt{\Gamma_i^{(k)} \over 2\pi}  \sqrt{1+\beta_i^{(k)} P_2(\cos\theta)} \; ,
\end{eqnarray}
up to a phase $\xi_i^{(k)}$, where $\Gamma_i^{(k)}$ is the partial rate of the core-excited state $k$ decaying into the final ion $i$, ${\bf \tilde{v}}_a$ is the velocity satisfying the energy conservation ${\bf \tilde{v}}_a^2/2 + E_i^{+} = E_k$, where $\theta$ is the angle between the x-ray polarization axis ${\bf e}_x$ and the velocity direction ${\bf v}_a$. $\beta_i^{(k)}$ is the anisotropy parameter that defines the angular distribution of the Auger electron emission and $P_2(\cos\theta)$ is the second Legendre polynomial. 

The core-excited state $1s^{-1}2p^{6}$ has a large number of Auger decay channels \cite{Krause1971,Albiez1990} and we only include in our model the channel with $2s^{-2}$($^1$S) as the final ion state. We focus on this decay since it is well separated in the Auger electron spectrum from the other channels and, thus, it allows us to clearly discern the sidebands.  The partial width is obtained from experiments \cite{Krause1971,Albiez1990}. Since the core-excited state, $1s^{-1}2p^{6}$, is not aligned, the Auger angular distribution is isotropic ($\beta_i^{(k)}$= 0) \cite{Cleff74,BookKabachnik}. See appendix \ref{app:2} for more details about the physical system.

Fig. \ref{Fig_1}(b) shows the angle-integrated Auger electron spectrum obtained with a peak x-ray intensity of $2.4\times10^{15}$ W/cm$^2$ and no optical laser field.  The x rays are linearly polarized with photon energy of 848 eV, which is in resonance with the transition $2p^{5} \leftrightarrow 1s^{-1}2p^{6}$. With this x-ray intensity we can transfer almost all the population from the ion ground state to the core-excited state, but it is not intense enough to produce Rabi oscillations between these two states. However, by increasing the peak intensity to $1.4\times10^{16}$ W/cm$^2$, we induce one and a half Rabi cycles (a $3\pi$ pulse for which the ion ground state is completely depleted, populated again, and completely depleted again) that modifies the Auger electron spectrum as studied in Ref. \cite{Nina2008}. Now, instead of having a single peak, we can clearly see the appearance of a twin-peak structure, as shown in Fig. \ref{Fig_1}(b).  This twin-peak structure is a result of the nonlinearity of the x-ray interaction via Rabi cycling.  The magnitude of the energy-level splitting between the two peaks is related to the AC Stark shift induced by the x rays in the core-excited state, as sketched in Fig. \ref{Fig_1}(a), but it is not exactly equal to the Rabi frequency $\Omega_x=\mu_{01}\varepsilon_{0x}$.  In particular, calculation of the magnitude of the energy-level splitting between the two peaks is more complicated and depends on more factors such as the decay of the core-excited state and the pulse shape of the x rays (see appendix \ref{app:3} for detailed discussion). 

If we now subject the ion to a strong optical field with a peak intensity of $10^{11}$ W/cm$^2$ in addition to an intense x-ray field with a peak intensity of  $1.4\times10^{16}$ W/cm$^2$, we see that sidebands appear in the Auger spectra, as shown in Fig. \ref{Fig_1}(c).   The presence of these sidebands is a result of the optical laser dressing of the continuum of the ion.  Note that the twin-peak structure induced by the strong x rays is preserved at individual sidebands separated by an energy equal to the optical photon energy. The sidebands have been labelled by the integer number $n$, where $n=0$ corresponds to the energy of the Auger decay channel (in our case 747.68 eV). The envelope of the sidebands presents a strong modulation when we calculate the Auger electron spectrum measured in the direction parallel to the optical polarization axis, defined as 0$^\circ$-angle. This modulation is called the gross structure of the sidebands \cite{Kazansky2010,Meyer2012}, which is the result of an interference between electron wavepackets emitted during one period of the optical field. We should remark that the multiphoton interferences presented in the next section have a different origin than the ones responsible for the gross structure. 


\section{Origin of the multiphoton interferences} \label{sec:origin}

\begin{figure}[b] 
\centerline{\includegraphics[width=0.75\columnwidth,clip]{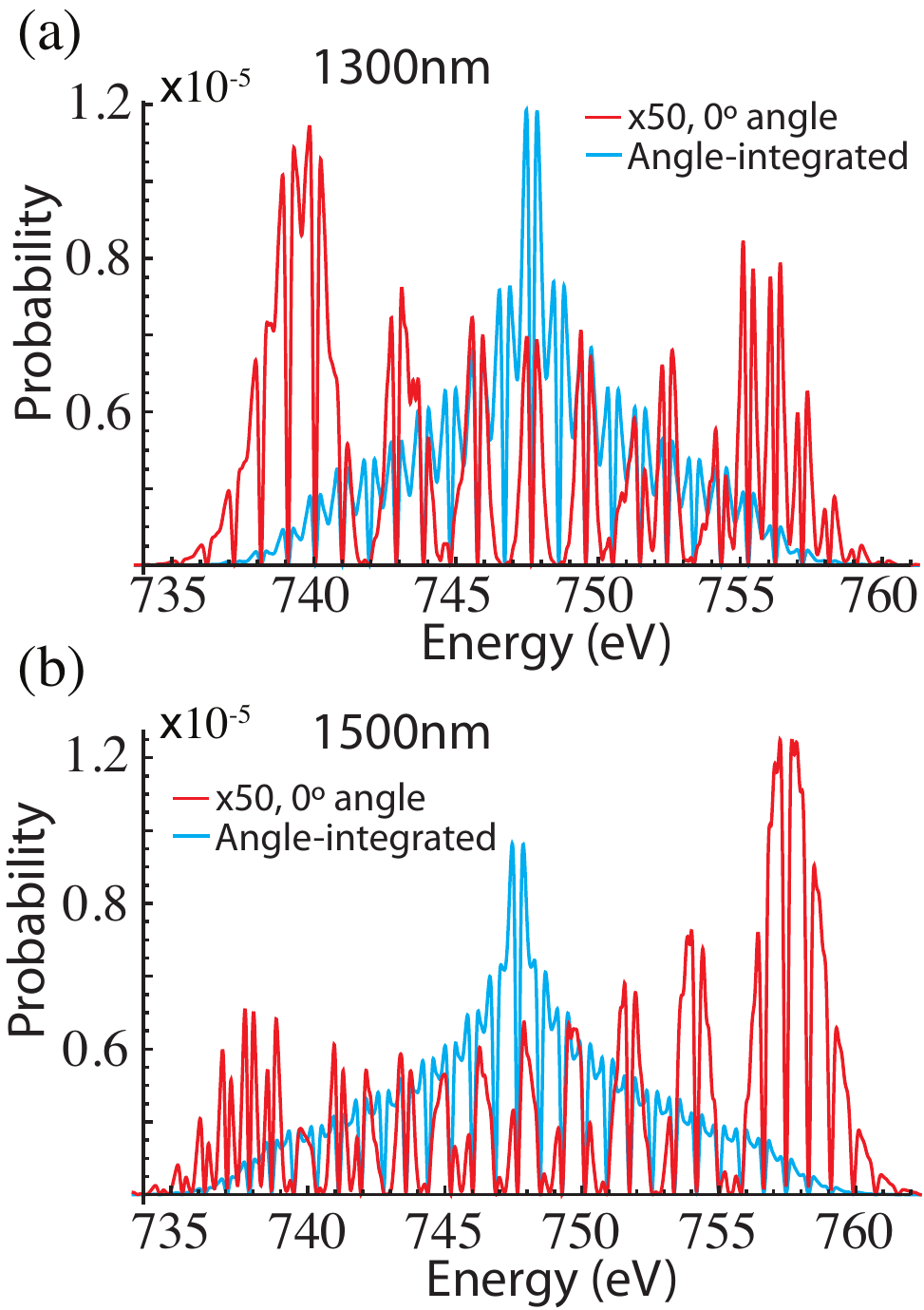}} 
\caption{(Color online)
The same as Fig. \ref{Fig_1}(c) but the optical wavelength is instead (a) 1300 nm and (b) 1500 nm.
}
\label{wavelength} \label{Fig_2}
\end{figure}


The multipeak structure generated by the sidebands and the strong x rays, as shown in Fig. \ref{Fig_1}(c), will display interference effects due to their coherence, and this is the main focus in the present work. By matching the optical frequency to the splitting of the Auger line induced by the intense x-rays, we can overlap the sidebands-peaks (e.g. the twin-peaks of $n=0$ with the twin-peaks of $n=1$) and expect strong interferences.  This overlap represents two possible quantum paths taken by the Auger electron; the two paths are from different quasi-energy levels induced by the strong x rays and from different optical transitions in the continuum due to the absorption or the emission of one more optical photons. In Fig. \ref{Fig_2} we show the Auger electron spectra obtained with a peak x-ray intensity of $1.4\times10^{16}$ W/cm$^2$.  But, in order to show the effect of the wavelength of the optical field, $\lambda_L$, we examine two cases calculated for $\lambda_L$ = 1300 nm and $\lambda_L$ = 1500 nm and compare them to the result for $\lambda_L$ = 800 nm, which is shown in Fig. \ref{Fig_1}(c). In these two cases the angle-integrated Auger electron spectra display a symmetric distribution centered at 747.68 eV, but the spectra emitted parallel to the optical polarization axis display strong asymmetries.  These asymmetries are evidences of the quantum-path interferences.  To quantify the degree of asymmetry, we use an asymmetry parameter defined as 
\begin{eqnarray}\label{Asymmetry}
A\equiv (n_a-n_b)/(n_a+n_b) \; ,
\end{eqnarray}
where $n_a$ is the probability of having the electron with energies above the energy of the Auger decay channel (747.68 eV) and $n_b$ is the probability of having the electron with energies below that.  We find that the 0$^\circ$-angle Auger electron spectra for $\lambda_L$ = 800 nm, 1300 nm and 1500 nm have $A$ =-0.02, -0.24 and 0.41, respectively.  The asymmetry is larger for 1500 nm as it increases the overlapping between sidebands of consecutive $n$, while the asymmetry is not appreciable for 800 nm as the overlapping is small. 

\begin{figure}[t] 
\centerline{\includegraphics[width=1.\columnwidth,clip]{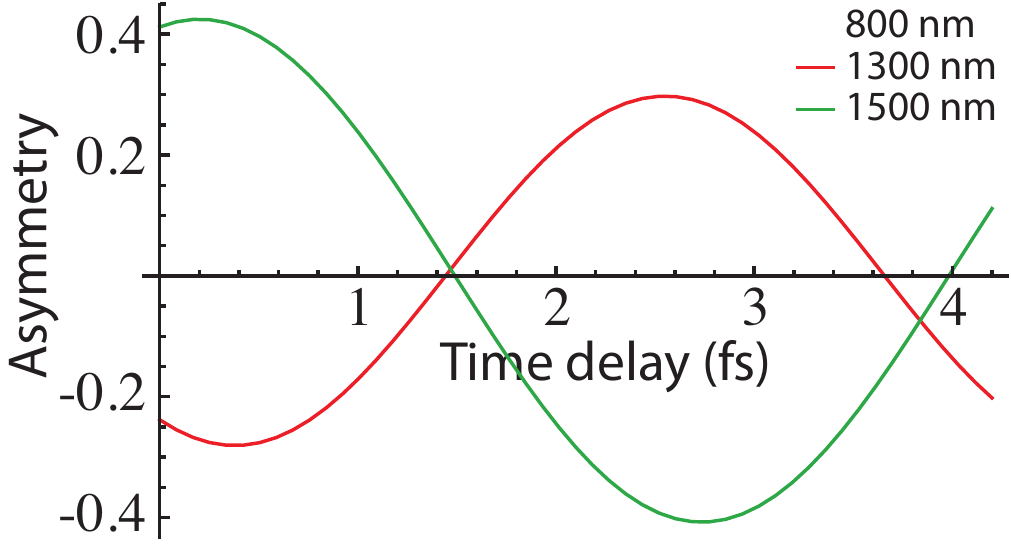}} 
\caption{(Color online)
Asymmetry parameter given by Eq. (\ref{Asymmetry}) with respect to the time delay between the optical field and the x-ray pulse.  We use the same parameters as in Fig. \ref{Fig_2} and only change the time delay of the x-ray pulse with respect to the optical field.
}
\label{wavelength} \label{Fig_3}
\end{figure}

In order to understand the origin of such coherent interferences, we examine an analytical solution of Eqs. (\ref{EOM}) for the case of treating the optical field as a continuous wave and the x rays as a square pulse (see appendix \ref{app:3}). The analytical solution obtained for this simple case allows us to qualitatively understand the asymmetries shown in Fig. \ref{Fig_2}. The interferences between the $n$ sidebands-peak and its consecutive neighbor peak are mainly described by the sum of a Bessel function $J_n$ times a function $I_{n}$, which is defined in Eq. (\ref{Integral}), specifically
\begin{eqnarray}
 i^{n} J_n\!\!\left({{\bf v}_{a}\cdot {\bf A}_0\over\omega_L}\right) e^{in\varphi}\; I_n (\Gamma_{1},\vert{\bf v}_{a}\vert,\varepsilon_{0x},t_f)+ \hspace{1.6cm} \nonumber\\
 i^{n+1} J_{n+1} \!\!\left({{\bf v}_{a}\cdot {\bf A}_0\over\omega_L}\right) e^{i(n+1)\varphi} \; I_{n+1}(\Gamma_{1},\vert{\bf v}_{a}\vert,\varepsilon_{0x},t_f) . \hspace{0.2cm} \label{Interferences} 
\end{eqnarray} 
This sum depends on several parameters, including the duration of the x rays ($t_f$), the peak electric field of the x rays ($\varepsilon_{0x}$), the frequency of the optical field ($\omega_L$), the optical vector potential (${\bf A}_0$), the decay rate of the core-excited state ($\Gamma_1$),  the carrier-envelope phase of the optical field ($\varphi$), and the velocity of the Auger electron (${\bf v}_{a}$). This sum then gives rise to the asymmetries observed in the Auger electron spectrum. 

Equation (\ref{Interferences}) reveals a unique feature, in which it shows that quantum interference can be controlled using the carrier-envelope phase of the optical field.  We exploit this effect of carrier envelope by varying the time delay between our Gaussian x-ray pulse and trapezoidal optical pulse.  Fig. \ref{Fig_3} shows that these asymmetries are very sensitive to the time delay between the optical field and the x-ray pulse, and the sensitivity is evident for all three optical frequencies, 800 nm, 1300 nm and 1500 nm.  The frequency of the modulation in $A$ is equal to the frequency of the optical field.  As expected, we observe stronger modulation in $A$ for $\lambda_L$ = 1500 nm and 1300 nm, while for $\lambda_L$ = 800 nm the asymmetry is small for all time delays.  

\begin{figure}[t] 
\centerline{\includegraphics[width=1.\columnwidth,clip]{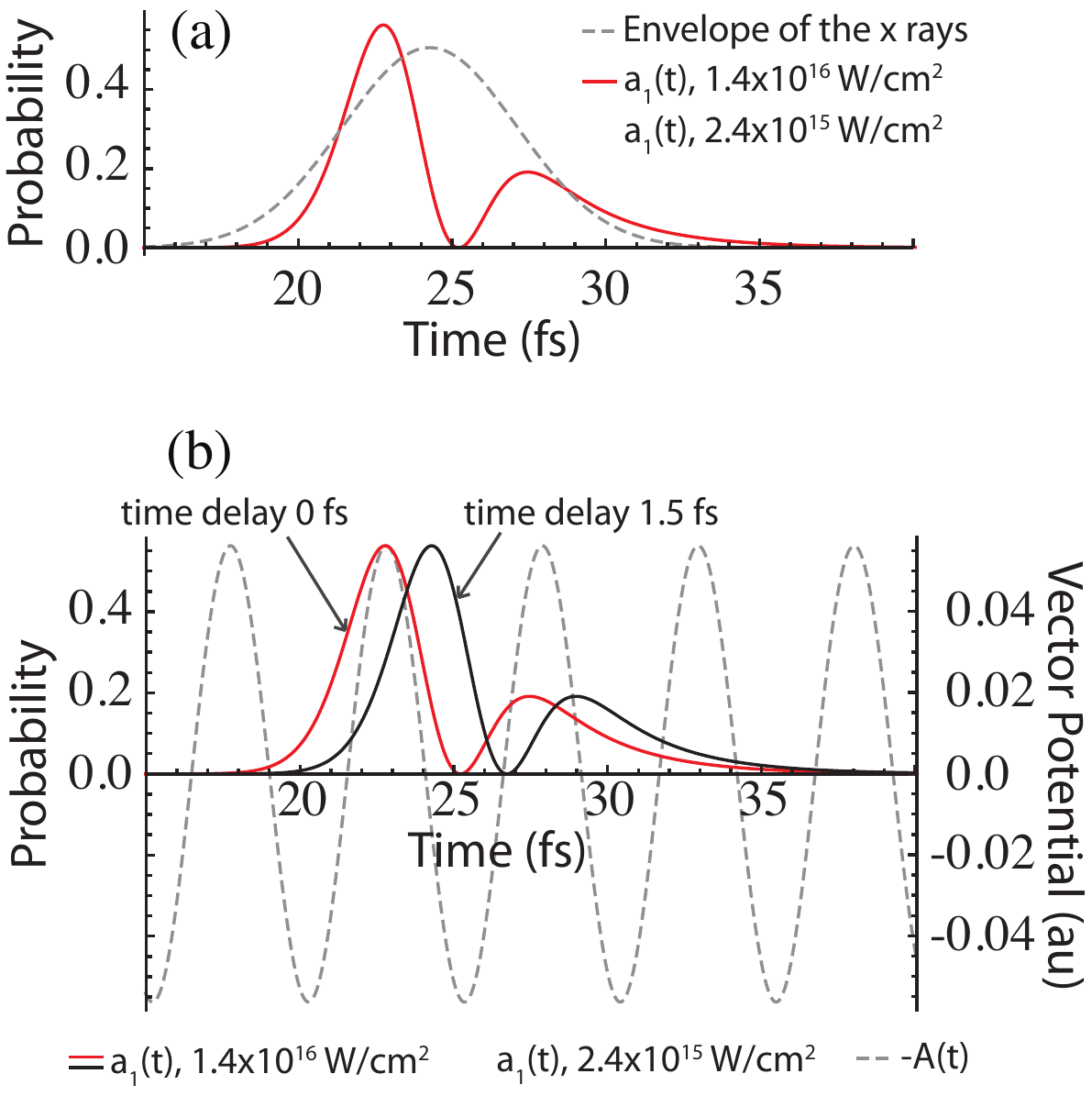}} 
\caption{(Color online)
(a) Evolution of the core-excited state population ($a_1(t)$) for two different x-ray intensities. We have used the same x-ray pulse as in Fig. \ref{Fig_1}(b). (b) The same as (a) but we compare the evolution of the core-excited state population with the optical vector potential with 1500 nm wavelength. For the case with $1.4\times10^{16}$ W/cm$^2$ intensity, we have considered a 0 fs and a 1.5 fs time delay between the x-ray pulse and the optical field.
}
\label{wavelength} \label{Fig_4}
\end{figure}

Although we have limited ourselves giving a physical interpretation in the energy domain picture, we can also provide a consistent description of the coherent multiphoton interferences in a time domain picture and connect the induced asymmetries with the well-known scenario of the optical streaking \cite{Itatani2002}. In optical streaking, the linear momentum of the emitted electron is shifted by the value of the vector potential at the emission time, therefore producing an asymmetry. In Fig. \ref{Fig_4}(a) we plot the evolution of the core-excited state population ($a_1(t)$) for the same x-ray pulse used in Fig. \ref{Fig_1}(b) for two different x-ray peak intensities.  In the case of the higher intensity of $1.4\times10^{16}$ W/cm$^2$, by matching the Rabi oscillation frequency to the optical frequency, we can obtain the scenario in which the electron population is mainly in the core-excited state only during the positive maxima of the optical vector potential.  As shown in Fig. \ref{Fig_4}(b), this scenario is realized when we use 0 fs time delay and $\lambda_L$ = 1500 nm. In this case we obtain a large asymmetry of A=0.41. However, for the lower intensity of $2.4\times10^{15}$ W/cm$^2$, the scenario is completely different, the core-excited state is populated for a wide range of vector potential amplitudes, which explains the lack of an asymmetry. It is noteworthy to relate this phenomenon with previous studies of optical streaking of Auger electrons \cite{Smirnova2003,Drescher2005,Kazansky2009,Kazansky2010}. In these previous works, it was observed that if the lifetime of the core- or inner-excited state is much shorter than half of the optical cycle, the Auger electron was streaked (energy shifted and therefore asymmetric) without showing the multipeak structure. When the lifetime is close to or larger than the optical period, the sidebands appear. This effect was explained by the interference of the electron wavepacket emitted at different times separated by an optical period.  Now, in our intense x-ray scenario, we can control the core-excited population using Rabi oscillation.  The Rabi cycling essentially shortens the core-excited lifetime and partially recovers the optical streaking scenario in which an energy shift is expected and so the asymmetry. Hence, by controlling the core-excited state population with the high-intensity x rays, we can restrict the timing of Auger decay to time windows when the amplitude of the vector potential is maximum and establish optical streaking at different times separated by the Rabi cycling period.


The time domain picture also explains the sensitivity of the asymmetries with respect to the carrier-envelope phase of the optical field. At 1500 nm, Fig. \ref{Fig_3} shows no asymmetry for a 1.5 fs time delay, and this is the scenario when the electron population is mainly in the core-excited state when the optical vector potential is almost zero, as shown in Fig. \ref{Fig_4}(b).  At further time delay of 2.5 fs, the excited state populations overlap the negative maxima of the vector potential, resulting in the maximum negative asymmetry.  

The fact that no interference effect is observed in the angle-integrated Auger electron spectrum can be understood by treating the light as photons. The Auger electron states with different $n$ correspond to a different number of absorbed/emitted photons during the Auger decay and they then have different angular momentum. Hence, the interference between the sideband-peaks $n$ and $n+1$ is analogous to the well-known scenario of the interferences of partial waves in atomic photoionization \cite{Kennedy1972}. Partial waves are incoherently added when the total photoionization cross section is calculated, however when the angle-resolved cross section is calculated they are coherently added.



\section{Observation of the asymmetries} \label{sec:observation}

Although the main goal of this work is to unveil the origin and the main effects of the coherent multiphoton interferences described in the last section, we would like to provide in this section a discussion of some factors that may play an important role for the observation of the asymmetries in the Auger electron spectrum. The first aspect is the transverse intensity distribution of the x rays, because not all the atoms in the gas cell or beam will experience the same intensity and then not undergo the same Rabi oscillations either. This naturally affects the asymmetry, as shown in Fig. \ref{Fig_5}(a). We observe that the asymmetry is strongly reduced for lower intensities, and can also change sign, as the broadening of the sideband peaks is narrower at lower intensities which reduces their overlaps. In this case, the contributions from the atoms interacting with the x rays at lower intensities may be too large to be able to measure any asymmetry at all. For example, we have considered an homogenous gas cell of Ne$^{+}$ and a transverse Gaussian distribution for the x rays with a peak intensity of $1.4\times10^{16}$ W/cm$^2$, the same intensity used in Fig. \ref{Fig_2}(b) (the width of the Gaussian distribution is not relevant since the gas cell is much larger than the transverse size of the beam). When we calculate the average asymmetry accounting for the intensity distribution, we obtain $A=0.05$ that is much smaller than the asymmetry $A=0.41$ given by the atoms in the peak intensity region. In the averaging we have also considered that at lower intensities less population from the ion ground state is transferred to the core-excited state. Therefore, in order to observe such asymmetries we need to prepare our gas target in a confined volume which is mainly affected by a constant intensity. For example, in a gas cell of Ne we can send a first pulse to ionize it and prepare Ne$^{+}$ only in the interacting area of the pulse that has the peak intensity. A similar scenario is considered in Refs. \cite{Kanter2011,Nina2012} where the same x-ray pulse that induces the Rabi oscillations also produces the Ne$^{+}$ gas target. 

The intensity averaging effects just described can be significant when reflective optics, such as Kirkpatrick-Baez mirrors, are used to achieve high peak intensities by focusing the x rays to small spot sizes with Gaussian transverse intensity distributions \cite{Linda2010,Gilles2011,Kanter2011}.  However, optical laser beams have been shaped to flat-top intensity profiles using refractive and diffractive optics and spatial modulators \cite{Singer1996,Huang1999,Bovatsek}.  Similar methods can be developed for x rays to produce flat-top intensities that will preserve large Auger asymmetries in experiments.

The wavelength of the optical field is another aspect that we need to take into account. As we discussed in the previous section, if the photon energy is large enough that there is no overlapping between the sideband-peaks, we do not observe any asymmetry in the electron spectrum. As the optical photon energy decreases, the asymmetry clearly increases. However, if the photon energy is too small, so that there is overlapping not only between the direct neighbor peaks, our qualitative description using Eq. (\ref{Interferences}) will obviously break down. The splitting of the twin-peaks induced by the strong x rays can be estimated as the generalized Rabi frequency, and therefore we can estimate that our analysis breaks down for photon energies below half that frequency, in our case below 0.018 a.u. (i.e. optical wavelengths longer than approximately 2500 nm). Naturally, if the optical wavelength is too long, we start to enter the optical streaking regime, where we observe an asymmetry even with low intensity x rays. 

\begin{figure}[t] 
\centerline{\includegraphics[width=1.\columnwidth,clip]{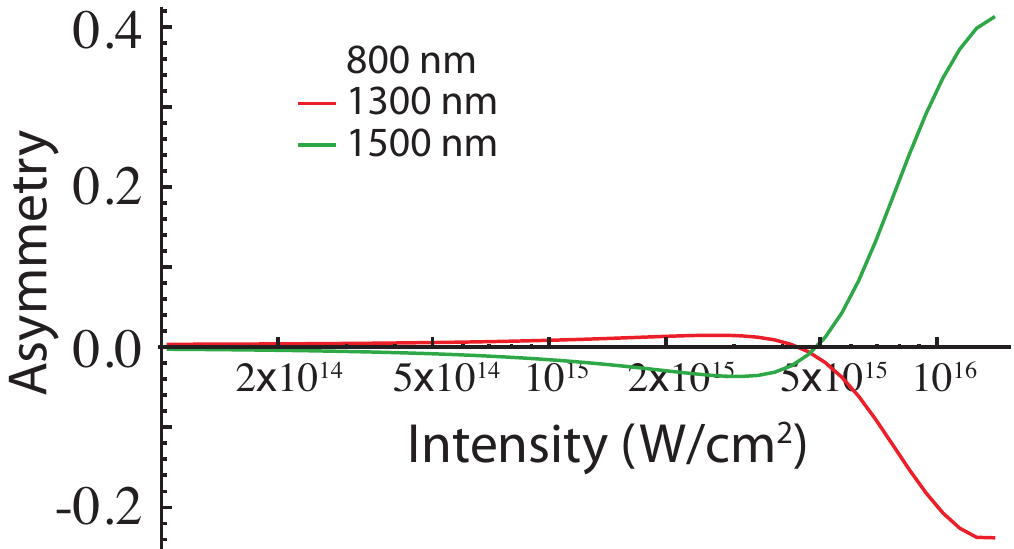}} 
\caption{(Color online)
Asymmetry parameter with respect to the intensity of the x rays. We used the same parameters as in Fig. \ref{Fig_2} and only changed the peak intensity of the x rays.
}
\label{wavelength} \label{Fig_5}
\end{figure}

Finally we would like to comment on the dependence of the asymmetry on the emission angle. In Fig. \ref{Fig_2} we only show the Auger electron spectrum measured at the angle $\theta=0$. At other angles, the envelope of the sidebands, the so-called gross structure \cite{Kazansky2010,Meyer2012}, is different, and in particular for $\theta=\pi/2$ (perpendicular to the polarization of the optical field), the sidebands vanish. The asymmetry remains approximately constant from $\theta=0$ to $\theta=\pi/2$, and it rapidly decreases to zero near $\theta=\pi/2$ when the sidebands vanish. After $\theta=\pi/2$, in the south hemisphere, the asymmetry is approximately the same as in the north hemisphere but with a change of sign, as the optical streaking acts now in the opposite direction. This behavior can be qualitatively explained using Fig. \ref{Fig_4}(b). The projection of the vector potential at different angles changes, however only its amplitude is modified, and the dynamics of the core-excited state population with respect to the vector potential is the same, giving rise to the same asymmetry. Hence, an angle-averaged measurement over a large angular range still presents the same asymmetry and allows the possibility of single shot measurements at fixed delay between the two pulses.




\section{Conclusions} \label{sec:conlusions}

In conclusion, we have analyzed resonant Auger processes under intense x rays and a strong optical field. We focus on the resonant transition $2p^{5} \!\leftrightarrow\! 1s^{-1}2p^{6}$ in Ne$^{+}$ decaying into the final ion $2s^{-2}$($^1$S). The optical field is strong enough to induce sidebands; a multipeak structure in which every sideband-peak is separated by an optical photon energy. If the x rays are intense enough, the sideband-peaks can be broadened and obtain a significant overlap between them. The overlap produces coherent interferences that originate from different multiphoton quantum paths taken by the Auger electron. We have shown that the interferences give rise to a clear signature: an asymmetry with respect to the energy and angle of Auger electron emission. Moreover, the interferences are very sensitive to the carrier-envelope phase of the optical field, i.e. the time-delay between the x-ray Gaussian pulse and the optical field. The intrinsic coherence of such interferences make them very sensitive to the x-ray and optical parameters, hence we believe that they can be exploited to develop alternative metrology tools for seeded pulses in FELs.

\begin{acknowledgments}
This work was supported by the Chemical Sciences, Geosciences, and Biosciences Division, Office of Basic Energy Sciences, Office of Science, U.S. Department of Energy, under Contract No. DE-AC02-06CH11357. We acknowledge fruitful discussions with Bertold Kr\"assig and Christian Buth. We thank Lahsen Assoufid for discussions of focusing optics.
\end{acknowledgments}

\appendix
\section{Ne$^{+}$ model} \label{app:2}

The main dynamics in Ne$^{+}$ is between the ground state ($\vert 0 \rangle$) and the core-excited state $1s^{-1}2p^{6}$ ($\vert 1 \rangle$), whose energies are obtained from experiments as $E_0=0$ eV and $E_1=848$ eV \cite{Kanter2011}.  The x rays couple the ion ground state with the $1s^{-1}2p^{6}$ state. The associated dipole moment is given by $\mu_{10} \!=\! \langle 1 \vert {\bf r}\cdot {\bf e}_x \vert 0 \rangle$, where ${\bf r}$ is the one-electron position operator and ${\bf e}_x$ is the polarization direction of the x rays. We obtain $\mu_{10}=0.0573$ a.u. by means of {\em ab initio} calculations \cite{Phay}.

We account for the x-ray photoionization and Auger decay rates in first-order of perturbation theory. Hence, the decay rate $\Gamma_k$ of the $k$ state depends on the non-radiative and radiative decay processes $\Gamma_{1s}$ as well as of the x-ray photoionization $\sigma_{k} J_{x}(t)$, i.e.
\begin{eqnarray}
\Gamma_k \!=\!\Gamma_{1s} (1-\delta_{0k})+ \sigma_{k} J_{x}(t) 
\end{eqnarray}
(for $k\!=\!0$ we do not have decay processes $\Gamma_{1s}$), where $\Gamma_{1s}$ is the natural linewidth of Ne$^+$ ($1s^{-1})$ (experimental value $\Gamma_{1s}\!=\!0.27$ eV, see references \cite{SchmidtBook,Coreno99}), $J_x(t)$ is the instantaneous x-ray flux \cite{McMorrowBook,ButhSantra07}, and $\sigma_{k}$ is the x-ray photoionization cross-section ($\sigma_0\!=\!2.56\times 10^{-20}$ cm$^2$ and $\sigma_1\!=\!3.47\times 10^{-20}$ cm$^2$, obtained using \cite{LosAlamos1} and \cite{LosAlamos2}). The intensity of the optical field is enough to dress the continuum but does not ionize the system.

The core-excited state may decay into a continuum state $\vert {\bf v}_a, i \rangle$, where $i$ represents the state of the final ion left after the decay with an Auger electron whose velocity is ${\bf v}_a$. Its energy is $E_i({\bf v}_a)={\bf v}_a^2/2 + E_i^{+}$, where ${\bf v}_a^2/2$ is the Auger electron kinetic energy and $E_i^{+}$ is the final ion energy. Under the strong-field approximation (SFA), we account for the continuum-continuum transitions responsible for sidebands, see more details in Ref. \cite{Picon2013}.

The core-excited state $1s^{-1}$ has a large number of decay channels (i.e. different final ion states $\vert i\rangle$) \cite{Krause1971,Albiez1990,Southworth2003}. Here we only consider the channel via Auger decay in which the final ion state is $2s^{-2}(^1S)$, whose energy is $E_{i}^{+}=100.32$ eV \cite{NIST}. This decay channel is analyzed because the main Auger peak is well separated in the spectrum from the other peaks. Therefore, even under the presence of the optical field, we can distinguish the sidebands from this channel from the other channels. The partial width is obtained from the experiment of Ref. \cite{Albiez1990} and is given by $\Gamma_{i}^{(1)} = 0.016$ eV . The core-excited state $1s^{-1}2p^{6}$ is not aligned, so the angular distribution is isotropic, i.e. $\beta_{i}^{(1)} = 0$ \cite{Cleff74,BookKabachnik}. 

We have not considered any effects of the optical field after Auger decay, such as Stark shift or ionization of the final ions, which are expected to be small. There are no resonant couplings between the considered final ion states, because they are out of resonance with the optical photon energy. We note that these final ion states are metastable and they present narrow linewidths in contrast to core-excited states.


\section{x rays and optical field} \label{app:1}
In all the numerical calculations, we use a Gaussian-envelope pulse for the x rays with electric field 
\begin{eqnarray}
\varepsilon_x (t) = \varepsilon_{0x} \exp[-(t\!-\!t_m)^2/2\sigma^2] \sin[\omega_X (t\!-\!t_m)] \; ,
\end{eqnarray}
where $\varepsilon_{0x}$ is the maximum amplitude of the electric field, $t_m$ is a given time when the electric field is maximum, $\sigma^2$ is the variance, and $\omega_X$ is the photon energy of the x rays. We use $\sigma=2$ fs, and this defines a bandwidth of $0.33$ eV, which is larger than the natural linewidth of the considered core-excited state. The small variance was chosen in order to induce Rabi oscillations before the characteristic time decay of the core-excited state. The photon energy is $\omega_X = 848$ eV, resonant with the ion ground state and the $1s^{-1}2p^{6}$ core-excited state transition.

We model the electric field of the optical laser as 
\begin{eqnarray} \label{EF}
\varepsilon_L (t)= f(t) \varepsilon_{0L} \cos[\omega_L t + \varphi] \; ,
\end{eqnarray}
where $f$($t$) is the field envelope function, $\varepsilon_{0L}$ is the maximum amplitude of the electric field, $\varphi$ is a phase that defines the continuous wave with respect to the arrival of the x rays, and $\omega_L$ is the photon energy of the optical field.  In order to ensure that both the electric field and the vector potential are zero at the beginning of each simulation, our field envelope function has a trapezoidal shape with 3-cycle turn-on, 3-cycle turn-off and 10-cycle plateau. We use a rather strong optical field with intensity $10^{11}$ W/cm$^2$.   Both x rays and optical field are linearly-polarized, and we use ${\bf e}_x$ and ${\bf e}_L$ to denote the polarization direction of the x-ray and optical laser pulse, respectively.  For our calculations, significant temporal overlap between the x-ray and the optical laser pulse takes place only during the plateau region of the optical pulse. The Auger electron spectrum is calculated at the end of the optical field.


\section{Analytical solution} \label{app:3}
The differential equation for the amplitudes of the continuum in Eq. (\ref{EOM}) can be expressed in the integral form as:
\begin{eqnarray} \nonumber
b_i ({\bf v}'_a,t) \!&=&\! -i \!\! \int_{t_0}^t \!\! dt'  \left\{ a_1 (t')  \gamma_{{\bf v}'_a+ {\bf A}(t'),1}   \right\} e^{-i\int_{t'}^{t}\! dt'' E_{{\bf v}'_{a} + {\bf A}(t'')}} \; , \\
\label{Amplitude_Auger_Electron}
\end{eqnarray}
where
\begin{eqnarray*}
&&{\bf v}'_a = {\bf v}'_a({\bf v}_a,t) = {\bf v}_a - {\bf A} (t) \; ,\\
&&E_{{\bf v}'_{a} + {\bf A}(t'')} = {\left({\bf v}'_{a} + {\bf A}(t'') \right)^2\over 2} + E_i^{+} \; .
\end{eqnarray*}
${\bf v}'_a$ is a time-dependent streaked velocity that depends on the vector potential of the optical field. Note that when the optical field vanishes, both velocities are the same, i.e. ${\bf v}'_a = {\bf v}_a$. By using Eq. (\ref{Auger_moment}), we can recast Eq. (\ref{Amplitude_Auger_Electron}) as
\begin{eqnarray} \nonumber
b_i ({\bf v}'_a,t) \!\sim\! -i {e^{i\xi_i^{(1)}} \over\sqrt{(4\pi \vert {\bf \tilde{v}}_a\vert)}} \sqrt{\Gamma_i^{(1)} \over 2\pi}  \sqrt{1+\beta_i^{(1)} P_2(\cos\theta)} \hspace{1cm} \\
 \times \!\! \int_{t_0}^t \!\! dt' a_1 (t') e^{-i\int_{t'}^{t}\! dt'' E_{{\bf v}'_{a} + {\bf A}(t'')}}  \; . \hspace{1cm}
\label{Amplitude_Auger_Electron_2}
\end{eqnarray}
From equation (\ref{Amplitude_Auger_Electron_2}) one can see that the excitation of the continuum depends on the core-excited amplitude dynamics as well as the optical field that changes the time-dependent phase in the integral. We can further simplify Eq. (\ref{Amplitude_Auger_Electron_2}) by assuming that the optical field is a continuous wave as Eq. (\ref{EF}) and neglecting the square terms of the vector potential with respect to the linear terms, we obtain
\begin{eqnarray} \label{Amplitude_Auger_Electron_CW}
b_i ({\bf v}'_a,t) \!\sim\! -i {e^{i\xi_i^{(1)}} \over\sqrt{(4\pi \vert {\bf \tilde{v}}_a\vert)}} \sqrt{\Gamma_i^{(1)} \over 2\pi}  \sqrt{1+\beta_i^{(1)} P_2(\cos\theta)} \hspace{1cm} \nonumber \\
\times \; e^{-i {{\bf v}'^2_{a}\over2} t} e^{-iE_i^+t} e^{-i{{\bf v}'_{a}\cdot {\bf A}_0\over\omega_L}\cos(\omega_L t + \varphi)} \hspace{1cm} \nonumber \\
\times \!\!\!\! \sum_{n=-\infty}^{\infty} \!\! i^{n} J_n\!\!\left({{\bf v}'_{a}\cdot {\bf A}_0\over\omega_L}\right) e^{in\varphi}
\!\! \int_{t_0}^t \!\! dt' c_1 (t') \; e^{i\left(\!{{\bf v}'^2_{a}\over2} + E_i^+ - E_1 + n\omega_L \!\! \right) t'} \!, \nonumber \\
\end{eqnarray}
where 
\begin{eqnarray*}
&& a_1(t)=c_1(t)e^{-iE_1 t}\; , \\
 && \varepsilon_{0L}  = \omega_L A_0 \; .
\end{eqnarray*}
From Eq. (\ref{Amplitude_Auger_Electron_CW}) we can understand the multipeak structure of the sidebands. The integral (\ref{Amplitude_Auger_Electron_CW}) is like a Fourier transform of the core-excited amplitude dynamics. The Fourier structure given by the integral is repeated in the Auger electron energy at every optical frequency, due to the term $n \omega_L$. Now, these repeated peaks are modified by the envelope caused by the Bessel function $J_n$. If the Fourier transform of the integral forms a localized energy function that does not overlap the sideband peaks and ${\bf v}'^2_{a}/2\gg \omega_L$, we obtain a symmetric multipeak structure with respect to $n=0$ due to the relation $J_{-n}(x) = (-1)^{n}J_n(x)$.

If the x rays are also a continuous wave, we can find an analytical solution for the core-excited amplitude \cite{Meystre_Sargent}:
\begin{eqnarray*}
c_1(t)= i \,{\Omega_x\over\Omega_{10}} e^{-{\Gamma_{1}\over4}t} \; \sin{\left(\Omega_{10} t\over2\right)} \; ,
\end{eqnarray*}
where the Rabi frequency is defined as $\Omega_x=\mu_{10} \varepsilon_{0x}$ and the generalized Rabi frequency as $\Omega_{10} = \sqrt{\Omega_x^2-(\Gamma_{1}/2)^2}$. If we have a square pulse from $t=0$ to $t=t_f$, the integral in Eq. (\ref{Amplitude_Auger_Electron_CW}) can be performed:

\begin{widetext}
\begin{eqnarray}
I_n &\equiv& i \,{\Omega_x\over\Omega_{10}} \int_0^{t_f} \!\! dt' \left[e^{-{\Gamma_{1}\over4}t'}  \sin{\Omega_{10} t'\over2} \right] e^{i \alpha_n t'} \nonumber \\
&& + i \,{\Omega_x\over\Omega_{10}} \left[e^{-{\Gamma_{1}\over4}t_{f}}  \sin{\Omega_{10} t_f \over2} \right]  \int_{t_f}^{\infty} \!\! dt' e^{-{\Gamma_{1}\over2}(t'-t_f)} e^{i\alpha_n t'} \nonumber \\
&& \equiv I_n^{(1)} + I_n^{(2)}\; , \label{Integral}
\end{eqnarray} 
where $\alpha_n= {{\bf v}'^2_{a}\over2} + E_i^+ - E_1 + n\omega$, and
\begin{eqnarray}
I_n^{(1)} =  i \,{\Omega_x\over\Omega_{10}} \left[e^{-{\Gamma_{1}\over4} t_f} e^{i\alpha_n t_f} {{\Omega_{10}  \over2} \cos ({\Omega_{10} \over2}\,t_f) + ({\Gamma_{1}\over4}-i\alpha_n)\sin({\Omega_{10}  \over2}\,t_f)\over (\alpha_n +i{\Gamma_{1}\over4} - {\Omega_{10}  \over2})(\alpha_n +i{\Gamma_{1}\over4}+ {\Omega_{10}  \over2})} \right. \nonumber \\ \left.
- {{\Omega_{10} \over2} \over (\alpha_n +i{\Gamma_{1}\over4}- {\Omega_{10}  \over2})(\alpha_n +i{\Gamma_{1}\over4}+ {\Omega_{10}  \over2})} \right] \; , \label{Integral1} \\
I_n^{(2)} = -i \,{\Omega_x\over\Omega_{10}} \left[e^{-{\Gamma_{1}\over4}t_{f}}  \sin{\left(\Omega_{10} t_f \over2\right)} \right] {e^{i\alpha_n t_f}\over -{\Gamma_{1}\over2} + i\alpha_n} \nonumber =\\ 
- \,{\Omega_x\over\Omega_{10}} \left[e^{-{\Gamma_{1}\over4}t_{f}}  \sin{\left(\Omega_{10} t_f \over2\right)} \right] {e^{i\alpha_n t_f}\over \alpha_n +i{\Gamma_{1}\over2}} \; . \label{Integral2}
\end{eqnarray} 
\end{widetext}

The solution of Eq. (\ref{Integral}) provides an insight to the Auger electron spectrum. We can start by analyzing the limit when the time duration of the square pulse is long compared with the lifetime of the core-excited state ($t_f \gg 1/\Gamma_1$). In that case only the second term of $I_n^{(1)}$ (\ref{Integral1}) survives. When the intensity of the x rays is large enough ($\Omega_{10}\gg\Gamma_1$), that term can be interpreted as two Lorentzian functions centered at electron energies satisfying $\alpha_n\pm\Omega_{10}/2=0$, with broadening $\Gamma_1/2$. Therefore, as the intensity increases, the separation between the two Lorentzian peaks increases. This effect is reminiscent of the Autler-Townes splitting. The first term of $I_n^{(1)}$, which is related to the finite time-length of the pulse, has the same denominator than the second term of $I_n^{(1)}$, giving rise thus to a two-Lorentzian-like profile. On the other hand, the term in $I_n^{(2)}$ (\ref{Integral2}) is related to the finite duration of the pulse, but at variance with the first term in $I_n^{(1)}$, the denominator gives rise to only one Lorentzian function centered at the electron energy satisfying $\alpha_n=0$, with broadening $\Gamma_1$. The numerators of the first term of $I_n^{(1)}$ and the term of $I_n^{(2)}$ account for the effects due to the finite time-length of the pulse, which are important for describing the Auger electron spectrum induced by Rabi cycling \cite{Nina2008}.

The integral (\ref{Integral}) allows us to describe qualitatively the interferences between peaks of different $n$. Introducing Eq. (\ref{Integral}) into Eq. (\ref{Amplitude_Auger_Electron_CW}), one can see that if the x rays are intense enough and the optical frequency is small enough, there is a strong interference between the functions
\begin{eqnarray}
 i^{n} J_n\!\!\left({{\bf v}'_{a}\cdot {\bf A}_0\over\omega_L}\right) e^{in\varphi} I_n + \hspace{2cm} \nonumber\\
 i^{n+1} J_{n+1} \!\!\left({{\bf v}'_{a}\cdot {\bf A}_0\over\omega_L}\right) e^{i(n+1)\varphi} I_{n+1} \; . \label{Interferences_2} 
\end{eqnarray} 
These interferences are very complex, in particular if one accounts for the finite time-length effects. However, there is a particular feature of Eq. (\ref{Interferences_2}) that is quite remarkable and it is the main result exploited in this work.  We see that the expression in Eq. (\ref{Interferences_2}) depends strongly on $\varphi$. 



\begin{thebibliography}{99}
%
\bibitem{Linda2010}
L. Young {\em et al.}, Nature {\bf 466}, 56 (2010).
%
\bibitem{Nina2007}
Nina Rohringer and Robin Santra, Phys. Rev. A {\bf 76}, 033416 (2007).
%
\bibitem{Fritzsche2009}
S. Fritzsche, A.N. Grum-Grzhimailo, E.V. Gryzlova, and N.M. Kabachnik, J. Phys. B: At. Mol. Opt. Phys. {\bf 42}, 145602 (2009).
%
\bibitem{Makris2009}
M.G. Makris, P. Lambropoulos, and A. Miheli\v{c}, Phys. Rev. Lett. {\bf 102}, 033002 (2009).
%
\bibitem{Meyer2010}
M. Meyer {\em et al.}, Phys. Rev. Lett. {\bf 104}, 213001 (2010).
%
\bibitem{Gilles2011}
G. Doumy {\em et al.}, Phys. Rev. Lett. {\bf 106}, 083002 (2011).
%
\bibitem{Hishikawa2011}
A. Hishikawa {\em et al.}, Phys. Rev. Lett. {\bf 107}, 243003 (2011).
%
\bibitem{Nina2008}
Nina Rohringer and Robin Santra, Phys. Rev. A {\bf 77}, 053404 (2008).
%
\bibitem{Kanter2011}
E.P. Kanter {\em et al.}, Phys. Rev. Lett. {\bf 107}, 233001 (2011).
%
\bibitem{Nina2012}
Nina Rohringer and Robin Santra, Phys. Rev. A {\bf 86}, 043434 (2012).
%
\bibitem{Demekhin2011}
Philipp V. Demekhin and Lorenz S. Cederbaum, Phys. Rev. A {\bf 83}, 023422 (2011).
%
\bibitem{Nikolopoulos2011}
L.A.A. Nikolopoulos, T.J. Kelly, and J.T. Costello, Phys. Rev. A {\bf 84}, 063419 (2011).
%
\bibitem{Demekhin2012}
Philipp V. Demekhin and Lorenz S. Cederbaum, Phys. Rev. Lett. {\bf 108}, 253001 (2012).
%
\bibitem{Cavaletto2012}
Stefano M. Cavaletto {\em et al.}, Phys. Rev. A {\bf 86}, 033402 (2012).
%
\bibitem{Amann2012} 
J. Amann {\em et al.}, Nature Photon. {\bf 6}, 693 (2012).
%
\bibitem{Allaria2012} 
E. Allaria {\em et al.}, Nature Photon. {\bf 6}, 699 (2012).
%
\bibitem{Drescher2002} 
M. Drescher {\em et al.}, Nature {\bf 419}, 803 (2002).
%
\bibitem{Predrag2011} 
P. Ranitovic {\em et al.}, Phys. Rev. Lett. {\bf 106}, 053002 (2011).
%
\bibitem{Bryan2012}
W.A. Bryan {\em et al.}, New J. Phys. {\bf 14}, 013057 (2012).
%
\bibitem{Glover2009}
T.E. Glover, M.P. Hertlein, S.H. Southworth, T.K. Allison, J. van Tilborg, E.P. Kanter, B. Kr\"assig, H.R. Varma, B. Rude, R. Santra, A. Belkacem, and L. Young, Nature Phys. {\bf 6}, 69 (2009).
%
\bibitem{Buth2007} 
C. Buth, R. Santra, and L. Young, Phys. Rev. Lett. {\bf 98}, 253001 (2007).
%
\bibitem{Dusterer2011} 
S. D\"usterer {\em et al.}, New J. Phys. {\bf 13}, 093024 (2011).
%
\bibitem{Schins1994} 
J.M. Schins {\em et al.}, Phys. Rev. Lett. {\bf 73}, 2180 (1994).
%
\bibitem{Smirnova2003} 
O. Smirnova, V.S. Yakovlev, and A. Scrinzi, Phys. Rev. Lett. {\bf 91}, 253001 (2003).
%
\bibitem{Drescher2005} 
M. Drescher and F. Krausz, J. Phys. B: At. Mol. Opt. Phys. {\bf 38}, S727 (2005).
%
\bibitem{Kazansky2009} 
A.K. Kazansky and N.M. Kabachnik, J. Phys. B: At. Mol. Opt. Phys. {\bf 42}, 121002 (2009).
%
\bibitem{Buth2009} 
C. Buth and K.J. Schafer, Phys. Rev. A {\bf 80}, 033410 (2009).
%
\bibitem{Kazansky2010} 
A.K. Kazansky and N.M. Kabachnik, J. Phys. B: At. Mol. Opt. Phys. {\bf 43}, 035601 (2010).
%
\bibitem{Meyer2012} 
M. Meyer {\em et al.}, Phys. Rev. Lett. {\bf 108}, 063007 (2012).
%
\bibitem{Agostini1979} 
P. Agostini et al., Phys. Rev. Lett. {\bf 42}, 1127 (1979). 
%
\bibitem{Picon2013} 
Antonio Pic\'on, Christian Buth, Gilles Doumy, Bertold Kr\"assig, Linda Young, and Stephen H. Southworth, Phys. Rev. A {\bf 87}, 013432 (2013).
%
\bibitem{ReviewMehlhorn}
W. Mehlhorn,  J. Electron Spectrosc. Relat. Phenom. {\bf 93}, 1-15 (1998).
%
\bibitem{ReviewArmen} 
G. Bradley Armen, Helena Aksela, Teijo \AA berg and Seppo Aksela, J. Phys. B: At. Mol. Opt. Phys. {\bf 33}, R49 (2000).
%
\bibitem{ReviewPiancastelli}
M.N. Piancastelli,  J. Electron Spectrosc. Relat. Phenom. {\bf 107}, 1-26 (2000).
%
\bibitem{Knight90} 
P.L. Knight, M.A. Lauder, and B.J. Dalton, Phys. Rep. {\bf 190}, 1 (1990).
%
\bibitem{Wigner30}
V. F. Weisskopf and E. P. Wigner, Z. Phys. {\bf 63}, 54 (1930).
%
\bibitem{Coreno99}
M. Coreno, L. Avaldi, R. Camilloni, K. C. Prince, M. de Simone, J. Karvonen, R. Colle, and S. Simonucci, Phys. Rev. A {\bf 59}, 2494 (1999).
%
\bibitem{SchmidtBook}
V. Schmidt, {\em Electron Spectrometry of Atoms Using Synchrotron Radiation} (Cambridge University Press, Cambridge, 1997).
%
\bibitem{Krause1971}
M. O. Krause, T. A. Carlson, and W. E. Moddeman, J. Phys. (Paris) {\bf32}, C4-139 (1971).
%
\bibitem{Albiez1990}
A. Albiez, M. Thoma, W. Weber, and W. Mehlhorn, Z. Phys. D: Atoms, Molecules and Clusters {\bf 16}, 97 (1990).
%
\bibitem{Cleff74}
B. Cleff and W. Mehlhorn, J. Phys. B: At. Mol. Phys. {\bf 7}, 593 (1974).
%
\bibitem{BookKabachnik} 
V.V. Balashov, A.N. Grum-Grzhimailo, and N.M. Kabachnik, {\em Polarization and Correlation Phenomena in Atomic Collisions: A Practical Theory Course} (Plenum Publishers, New York, 2000).
%
\bibitem{Itatani2002}
J. Itatani, F. Qu\'er\'e, G.L. Yudin, M.Yu. Ivanov, F. Krausz, and P.B. Corkum, Phys. Rev. Lett. {\bf 88}, 173903 (2002).
%
\bibitem{Kennedy1972}
D. J. Kennedy and S. T. Manson, Phys. Rev. A {\bf 5}, 227 (1972).
%
\bibitem{Singer1996}
W. Singer, H. P. Herzig, M. Kuittinen, and E. P. J. Wangler, Opt. Eng. {\bf 35}, 2779 (1996).
%
\bibitem{Huang1999}
X. G. Huang, M. R. Wang, and C. Yu, Opt. Eng. {\bf 38}, 208 (1999).
%
\bibitem{Bovatsek}
J. Bovatsek and R. S. Patel, ``High-power, nanosecond-pulse Q-switch laser technology with flattop beam-shaping technique for efficient industrial laser processing,'' http://www.newport.com/images/webdocuments-en/images/29436.pdf.


%
%
%
%
%
%
%
%
%
%
%
%
%
\bibitem{Phay}
P. J. Ho, ``XRAP, a Hartree-Fock-Slater atomic transition code''. 
%
\bibitem{McMorrowBook}
J. Als-Nielsen and D. McMorrow, {\em Elements of modern x-ray physics} (John Wiley \& Sons, New York, 2001).
%
\bibitem{ButhSantra07}
Christian Buth and Robin Santra, Phys. Rev. A {\bf 75}, 033412 (2007).
%
\bibitem{LosAlamos1}
Robert D. Cowan, {\em The theory of atomic structure and spectra}, Los Alamos Series in Basic and Applied Sciences (University of California Press, Berkeley, 1981) ISBN 9-780-520-03821-9.
%
\bibitem{LosAlamos2}
Los Alamos National Laboratory, Atomic Physics Codes, http://aphysics2.lanl.gov/tempweb/lanl/.
%
\bibitem{Southworth2003}
S.H. Southworth, E.P. Kanter, B. Kr\"assig, L. Young, G.B. Armen, J.C. Levin, D.L. Ederer, and M.H. Chen, Phys. Rev. A {\bf 67}, 062712 (2003).
%
\bibitem{NIST} A. Kramida, Yu. Ralchenko, J. Reader, and NIST ASD Team (2012). NIST Atomic Spectra Database (version 5.0), [Online]. Available: http://physics.nist.gov/asd. National Institute of Standards and Technology, Gaithersburg, MD.
%
\bibitem{Meystre_Sargent}
Pierre Meystre and Murray Sargent III, {\em Elements of quantum optics}, 3rd ed. (Springer, Berlin, 1999).

\end{thebibliography}
\end{document}